\documentclass[aps,prb,twocolumn,groupedaddress,showpacs]{revtex4}

\usepackage[bookmarks]{hyperref}
\usepackage[dvips]{graphicx}
\usepackage{amsmath}
\usepackage{amstext}
\usepackage{graphics}
\usepackage{exscale}
\usepackage{epsfig}
\usepackage[T1]{fontenc}
\usepackage{bm}
\usepackage{bbm}

\usepackage{latexsym}
\renewcommand{\epsilon}{\varepsilon}

\newcommand{\integral}[3]{\!\int\limits_{#2}^{#3}\!\!{\rm d}#1\;}

\newcommand{\elcre}[2]{ c^{\dagger}_{#1,#2}}
\newcommand{\elann}[2]{ c_{#1,#2}}

\newcommand{\thGf}{{\cal G}}

\newcommand{\hc}{\mathrm{h.c.}}

\begin{document}

\title{Quasiparticle properties of strongly correlated electron systems with
  itinerant metamagnetic behavior} 
\author{Johannes Bauer}
\affiliation{Max-Planck Institute for Solid State Research, Heisenbergstr.1,
  70569 Stuttgart, Germany}
%\affiliation{Department of Mathematics, Imperial College, London SW7 2AZ,
%  United Kingdom}
\date{\today} 
\begin{abstract}
A brief account of the zero temperature magnetic response of a system of
strongly correlated electrons in strong magnetic field is given in terms of
its quasiparticle properties. The scenario is based on the paramagnetic phase
of the half-filled Hubbard model, and the calculations are carried out with
the dynamical mean field theory (DMFT) together with the numerical
renormalization group (NRG). As well known, in a certain parameter regime one
finds a magnetic susceptibility which increases with the field strength. Here,
we analyze this metamagnetic response based on Fermi liquid parameters, which
can be calculated within the DMFT-NRG procedure. The results indicate that the
metamagnetic response can be driven by field-induced effective mass
enhancement. However, also the contribution due to quasiparticle interactions
can play a significant role. We put our results in context with experimental
studies of itinerant metamagnetic materials.
%heavy fermion materials.
%When compared with experimental studies of heavy fermion materials
%qualitatively consistent behavior for the field dependence of the
%susceptibility and effective mass is found.   
 \end{abstract}
%\PACS{ 72.15.Qm\sep 75.20Hr\sep 73.21.La}
%72.10.F Kondo transport
% 73.61 Quantum dots conductivity
% 71.10.A Fermi liquid theory
% 11.10.G Renormalization, field theory

\maketitle

\section{Introduction}
%\paragraph*{Introduction -}
The interplay of strong correlation physics and magnetic behavior in itinerant
electronic systems has been a fascinating subject for many years. At low
temperature it is often possible to describe the response of such systems in
terms of the low energy 
excitations and quasiparticle properties such as in a Fermi liquid picture. 
The ratio of the spin susceptibility of the interacting system $\chi_s$ and
that of the non-interacting system $\chi_s^0$ is then given by the expression
\begin{equation}
  \label{flsusc}
  \frac{\chi_s}{\chi_s^0}=\frac{m^*/m_0}{1+F_0^a},
\end{equation}
where $m^*/m_0$ is the ratio of effective and bare electronic mass, and $F_0^a$
is the lowest order asymmetric Landau parameter, which accounts for
quasiparticle interactions.  
A special kind of response is metamagnetism, which we define here as the
existence of a regime where the system's differential susceptibility
$\chi_s={\rm d}{M}/{\rm d} H$ increases with magnetic field $H$,
i.e. ${\rm d}\chi_s/{\rm d} H>0$, for $H\in[H_1,H_2]$ with $H_1>0$. 
The subject of this paper is the analysis of the metamagnetic response in
correlated electron systems in terms of the Fermi liquid
description (\ref{flsusc}). For this we calculate the effective mass and
the term due to quasiparticle interactions from a microscopic model. This
allows us to understand what drives the magnetic response. This can be relevant for the
interpretation of experiments for itinerant metamagnets where the
magnetic response is measured simultaneously with the field dependence of the
specific heat. 

In a naive single electron picture itinerant metamagnetism is not intuitive as with
increasing polarization the magnetic response usually decreases. 
For instance, in  weakly interacting systems, such as a Hubbard model with small $U$, with a
featureless concave density of states metamagnetic behavior does not occur. RPA
based calculations yield a decreasing susceptibility with increasing  
field as spin fluctuations are suppressed. 
%This suppression is often accompanied by a reduction in the effective masses.
On the other hand, a convex density of states, i.e. with positive curvature at the Fermi
energy, such as in the Wohlfahrt and Rhodes \cite{WR62} 
theory can lead to metamagnetic behavior. This is exploited in a number of works, where
the Hubbard model with such convex density of states is analyzed \cite{NH97,SO98}.
Metamagnetic behavior is shown to also occur in situations where the Fermi energy lies
close to a van Hove singularity \cite{BS04,Hon05}, or where a Pomeranchuk
Fermi surface deformation instability occurs\cite{YK07}.
It has been shown by calculations based on the Gutzwiller approximation by
Vollhardt\cite{Vol84} and Spalek and coworkers 
\cite{SG90,KSWA95,SKW97} that for a generic concave density
of states metamagnetic behavior is also found
in the intermediate coupling regime of the Hubbard model.
% induced by the electronic correlations only. 
The metamagnetic scenario is then that of correlated electrons, with a (Mott)
localization tendency due to the interaction.

Our calculations are based on the  half filled single band Hubbard model which has been
used  frequently to describe itinerant metamagnetism for correlated electrons 
\cite{LGK94,Tri95,KSWA95,SKW97,NH97,SO98,BS04,Hon05} due 
to its relative formal simplicity. We employ the dynamical
mean field theory (DMFT) \cite{LGK94,GKKR96} combined with the numerical
renormalization group (NRG) \cite{KWW80a,BCP08} to solve the effective
impurity problem.   
We focus on the case of zero temperature, where sharp features are most
clearly visible. We follow these earlier approaches here and restrict
ourselves to the response of the paramagnetic solutions of the Hubbard model,
which is possible for mean field-like approaches. 

The half filled Hubbard model in a magnetic field has
already been investigated by detailed DMFT studies by Laloux et
al. \cite{LGK94} and Bauer and Hewson \cite{BH07b}. Low temperature magnetization
curves and field induced metal insulator transitions have been 
investigated by Laloux et al. Metamagnetic response based on correlated electron
physics, seen in the Gutzwiller approach, was confirmed in such
calculations. Our analysis extends previous work\cite{LGK94} as we 
investigate the $T=0$ magnetic response with a Fermi liquid interpretation
based on the field dependent renormalized parameter
approach\cite{HOM04,BH07b,HBK06,BH07a}. This, together with results for the spectral 
functions, allows us to identify what gives rise to the magnetic response in the system.

The paper is organized as follows. In a brief section II we give details about
the model and method. The Fermi liquid interpretation and the relation between
Fermi liquid parameters and the field dependent renormalized parameters are
described in section III. Section IV reports the results for magnetization,
susceptibilities and the interpretation in terms of effective mass and
quasiparticle interactions.  We conclude by putting our results in context
with itinerant metamagnetism studied experimentally.
%heavy fermions systems.

\section{Model and Method}
%\paragraph*{Model and Background -}
The basis for our calculation forms the Hubbard Hamiltonian in a magnetic
field, which in the grand canonical formulation reads
\begin{equation}
H_{\mu}=\sum_{i,j,\sigma}(t_{ij}\elcre {i}{\sigma}\elann
{j}{\sigma}+\hc)-\sum_{i\sigma}\mu_{\sigma}
n_{i\sigma}+U\sum_in_{i,\uparrow}n_{i,\downarrow}\label{hubm}.
\end{equation}
$\elcre {i}{\sigma}$ creates an electron at site $i$ with spin $\sigma$, and
$n_{i,\sigma}=\elcre {i}{\sigma}\elann {i}{\sigma}$.  $t_{ij}=-t$ for nearest
neighbors is the hopping amplitude and $U$ is the on-site interaction;
$\mu_{\sigma}=\mu+\sigma h$, where $\mu$ is the chemical potential of the
interacting system, and the Zeeman splitting term  with external magnetic
field $H$ is given by $h=g\mu_{\rm B} H/2$ with the Bohr magneton $\mu_{\rm  B}$.  
%We are dealing with the one s-band Hubbard
%model here, so no coupling of the field to angular momentum states has to be included.
%From Dyson's equation, the Green's function $G_{{\vk},\sigma}(\omega)$ can
%be expressed in the form,  
%\begin{equation}
%G_{{\vk},\sigma}(\omega)
%=\frac{1}{\omega+\mu_\sigma-\Sigma_{\sigma}({\vk},\omega)-\epsilon({\vk})},  
%\end{equation}
%where $\Sigma_{\sigma}({\vk},\omega)$ is the proper self-energy,
%and $\epsilon({\vk})=\sum_{\vk}e^{-{\vk}\cdot({\vct R}_i-{\vct R}_ j)}t_{ij}$.
In the DMFT approach the proper self-energy is a function of $\omega$ only 
\cite{MV89,Mue89}.
%, which is exact for the models in the infinite dimensional limit. 
In this case the local lattice Green's function    
$ G_{\sigma}^{\mathrm{loc}}(\omega)$ can be expressed in the form,
\begin{equation}
G_{\sigma}^{\mathrm{loc}}(\omega) =
%\sum_{\vk}G_{{\vk},\sigma}(\omega)=
\integral{\epsilon}{}{}\frac{\rho_0(\epsilon)}
{\omega+\mu_\sigma -\Sigma_{\sigma}(\omega)-\epsilon},
\label{gloc} 
\end{equation}
where $\rho_0(\epsilon)$ is the density of states for the non-interacting model
($U=0$). 
It is  possible to convert this lattice problem into an effective
impurity one \cite{GKKR96}, introduce the dynamical Weiss field
$\thGf_{0,\sigma}^{-1}(\omega)$. The DMFT self-consistency condition reads
\begin{equation}
  \thGf_{0,\sigma}^{-1}(\omega)=G_{\sigma}^{\mathrm{loc}}(\omega)^{-1}
  +\Sigma_{\sigma}(\omega).
\label{tgf}
\end{equation}
The Green's function  $ G_{\sigma}^{\mathrm{loc}}(\omega)$ can  be
identified with the Green's function $ G_{\sigma}(\omega)$ of an effective
Anderson model,  and $\thGf_{0,\sigma}^{-1}(\omega)$ expressed as
\begin{equation}
\thGf_{0,\sigma}^{-1}(\omega)=\omega+\mu_{\sigma}-K_{\sigma}(\omega).
\label{thgfK}
\end{equation}
The function $K_\sigma(\omega)$ plays the role of a dynamical mean field 
describing the effective medium surrounding the impurity. 
% In the impurity case in the wide band limit we have
% $K_{\sigma}(\omega)=-i\Delta$. 
$K_\sigma(\omega)$ and $\Sigma_\sigma(\omega)$  have to be calculated
self-consistently using equations (\ref{gloc})-(\ref{thgfK}). Our calculations
are based on the numerical NRG\cite{KWW80a,BCP08} to solve the effective 
impurity problem.  As in earlier work\cite{BH07b} we calculate spectral
functions from a complete basis 
set\cite{PPA06,WD07} and use higher Green's functions to obtain the
self-energy \cite{BHP98}. For numerical calculations within the DMFT-NRG
approach for $\rho_0(\epsilon)$ we take the semi-elliptical form for the
non-interacting density of states  
%\begin{equation}
 $\rho^{\rm sem}_0(\epsilon)={2}\sqrt{D^2-\epsilon^2}/{\pi D^2}$,
%\label{dos}
% \end{equation}
where $W=2D$ is the band width with $D=2t$ for the Hubbard model. $t=1$ sets
the energy scale in the following. 

\section{Field dependent renormalized parameters and Fermi liquid theory}
% Fermi liquid theory here
The response of a metallic system of correlated electrons can often be described in
terms of Fermi liquid theory. The ratio of the spin susceptibility 
of the interacting system $\chi_s$ and that of the non-interacting system
$\chi_s^0$ is given in equation (\ref{flsusc}).
Thus, when strongly interacting fermions have a large paramagnetic
susceptibility, it can be interpreted as due to quasiparticles with
large effective masses. It is, however, also possible that the susceptibility
is additionally enhanced due to the quasiparticle interaction term $1/[1+F_0^a]$, which
is for instance the case in liquid ${}^3\rm He$, where $m^*/m_0\simeq 5$ but
$\chi_s/\chi_s^0\simeq 20$.\cite{BFSWRPW98} This is usually described by the
dimensionless Sommerfeld or Wilson  ratio $R$ of the magnetic susceptibility
and the linear specific heat coefficient $\gamma$. We will use it in the form
$R=({\chi_s}/{\chi_s^0})/(\gamma/\gamma_0)$, where $\gamma/\gamma_0=m^*/m_0$. 

Here we are interested in analyzing the behavior in finite field, and it is
possible to calculate corrections of higher order in $H$ to equation
(\ref{flsusc}).\cite{Mis71}  We will, however, follow a different approach here, and
assume that expression (\ref{flsusc}) remains valid for finite field with
field dependent effective mass $m^*(H)$ and Landau parameter $F_0^a(H)$. This
is in the spirit of the field dependent quasiparticle parameters
introduced in earlier work \cite{BH07b,HBK06,BH07a}. 
Notice that for the case considered the field dependence of $\chi_s^0$, which is
given by the non-interacting density of states, varies very little in the relevant field
range. In this picture with field dependent parameters, metamagnetism can occur
when the  effective mass increases with the magnetic field. Generally,
however, also the field dependence of the quasiparticle interaction plays a
role. One hypothesis, tested in this paper, is that itinerant metamagnetic
behavior is always  accompanied by a field induced localization and a sharp
increase of the effective mass near the metamagnetic transition. 
%Such a behavior was found based on the Gutzwiller approximation \cite{Vol84} and the FLEX approach
%\cite{JC00}.   
%Metamagnetism is also found in calculations based on the Gutzwiller approach
%\cite{Vol84,SG90,KSWA95}. Compared with DMFT results it is shown by Laloux et
%al. \cite{LGK94} that the occurrence of metamagnetic behavior is overestimated
%in the Gutzwiller approach and the values for the quasiparticle weight were
%found to be significantly smaller than the Gutzwiller predictions, the ratio is more 
%than a factor of $2$ for $U>4$ and zero field. 

%These large effective masses arise 
%from the scattering of the electrons with the enhanced spin fluctuations
%induced by the strong local Coulomb interactions. 

%Without these
%features a normal paramagnetic response is found in such systems, and, as we
%will see, metamagnetic response occurs when strong correlation effects are
%taken into account properly.

In order to calculate the microscopic Fermi liquid parameters, we expand
$\Sigma_\sigma(\omega)$ in powers of $\omega$ for small $\omega$, and retain
terms to first order in $\omega$ only. This is used to define 
renormalized parameters\cite{BH07b}
\begin{equation} 
\tilde\mu_{0,\sigma}=z_{\sigma}[\mu_{\sigma}-\Sigma_{\sigma}(0)],\quad{\rm and}\quad
     z_{\sigma}=1/[1-\Sigma'_{\sigma}(0)].
\label{nrgqp}
\end{equation} 
and from (\ref{gloc}) a normalized quasiparticle propagator,  
%$\tilde G_{0,\sigma}^{\mathrm{loc}}(\omega)$, 
\begin{equation}
\tilde G_{0,\sigma}^{\mathrm{loc}}(\omega)
=\frac1{z_{\sigma}}\integral{\epsilon}{}{}\frac{\rho_0(\epsilon/z_{\sigma})} 
{\omega+\tilde\mu_{0,\sigma} -\epsilon}.
\label{gqploc}
\end{equation}
Note that this $\omega$-expansion can also be carried out in finite magnetic
field. Then the renormalized parameters become field dependent,
$z_{\sigma}=z_{\sigma}(h)$ and $\tilde\mu_{0,\sigma}=\tilde\mu_{0,\sigma}(h)$.
The density of states  $\tilde \rho_{0,\sigma}(\epsilon)$
derived from (\ref{gqploc}), $\tilde \rho_{0,\sigma}(\epsilon)=-{\rm Im}\tilde
G_{0,\sigma}(\epsilon+i\delta)/\pi=\rho_0[(\epsilon+\tilde\mu_{0,\sigma})/z_{\sigma}]/z_{\sigma}$,
is referred to as the free quasiparticle density of states. $z_\sigma$
is interpreted as the weight of the quasiparticle resonance and $\tilde\mu_{0,\sigma}$ 
gives the position of the quasiparticle band. All energies are
measured from the chemical potential $\mu$.

To obtain the renormalized parameters $z_\sigma$ and
$\tilde\mu_{0,\sigma}$, we use two different methods based on the NRG
approach. The first method is a direct one where we use the 
self-energy $\Sigma_\sigma(\omega)$ determined by NRG and the chemical
potential  $\mu_\sigma$, and then substitute into  equation (\ref{nrgqp}) for
$z_\sigma$ and $\tilde\mu_{0,\sigma}$. The second method is indirect,  and
it is based on the quasiparticle interpretation of the NRG low energy fixed
point of the effective impurity.\cite{HOM04} This approach has been used
earlier for the Hubbard model \cite{BH07b,BH07c} and for the Anderson impurity
model in a magnetic field \cite{HBK06,BH07a}. As shown before the results of
both methods usually agree within a few percent, and we use an average value of both
methods for the numerical results presented later. It is important to
calculate these parameters accurately, since for the following results also
their derivatives are needed. 
%It is therefore useful to have two methods at
%hand to calculate the parameters, especially since there can occur numerical
%problems to calculate derivative of the self-energy due to the NRG broadening
%procedure. 

We can calculate static expectation values and response functions in terms
of the renormalized parameters. The quasiparticle occupation number $ \tilde
n^0_{\sigma}$ is given by integrating the quasiparticle density of states up
to the Fermi level,     
 \begin{equation}
   \tilde n^0_{\sigma}=\integral{\epsilon}{-\infty}{0}\tilde
   \rho_{0,\sigma}(\epsilon)=\integral{\epsilon}{-\infty}{\infty}\rho_{0,\sigma}(\epsilon)
   \theta(\mu_{\sigma}-\Sigma_{\sigma}-\epsilon).
\label{qpocc}
 \end{equation}
Luttinger's theorem \cite{Lut60} holds for each spin
component for the Hubbard model in magnetic field\cite{BH07b}, hence we have
$\tilde  n^0_{\sigma}= n_{\sigma}$, where  $n_{\sigma}$ is the value of the
occupation number in the interacting system at $T=0$.

To calculate the magnetic response we focus for the rest of this paper on the
case with particle-hole symmetry where $\mu=U/2$, and we can 
write $\Sigma_{\sigma}(0,h)=U/2-\sigma\eta(h)$. 
We can calculate $\eta(h)$ directly from the self-energy, 
e.g. $\eta(h)=(\Sigma_{\downarrow}- \Sigma_{\uparrow})/2$, or from the renormalized
parameters $\eta(h)=\tilde\mu_0(h)/z(h)-h$. At half filling we have
$z_{\uparrow}=z_{\downarrow}\equiv z$ and
$\tilde\mu_{0,\uparrow}=-\tilde\mu_{0,\downarrow}\equiv \tilde\mu_{0}$. We
define the function 
\begin{equation}
g(h):=h+\eta(h)=\tilde\mu_0(h)/z(h)=\tilde\mu_0(h)m^*(h)/m_0, 
\end{equation}
as $m^*/m_0=z^{-1}$ in DMFT. In terms of the quasiparticles it is the product
of the effective mass enhancement $m^*/m_0$ and the shift of the quasiparticle band
$\tilde\mu_0$. With the applicability of Luttinger's theorem the magnetization 
is then given by  
\begin{equation}
  \label{mag}
  m(h)=\frac12(n_{\uparrow}-n_{\downarrow})=\integral{\epsilon}{-\infty}{\infty}
  \rho_0(\epsilon)\theta[g(h)-\epsilon] -\frac12.
\end{equation}
For a local self-energy this is an exact expression for the magnetization, which only 
depends on the field dependent renormalized parameters via $g(h)$.  
For certain bare densities of state, for instance, for the semi-elliptical
density of states $\rho^{\rm sem}_0(\epsilon)$, it can be evaluated analytically,
\begin{equation}
   m(h)=
\frac12 g(h)\rho^{\rm sem}_0(g(h))+\frac1{\pi}\arcsin(g(h)).
  \label{mag2}
%\frac12 (\tilde\mu_0\frac{m^*}{m_0})\rho_0(\tilde\mu_0\frac{m^*}{m_0})+\frac1{\pi}\arcsin(\tilde\mu_0\frac{m^*}{m_0})
\end{equation}
Differentiating (\ref{mag}) with respect to $h$ yields the local static spin susceptibility 
\begin{equation}
  \label{suscrp}
  \chi_s=\frac{{\rm d}m}{{\rm d}h} =g'(h)\rho_0(g(h))
%=\partdera{h}{(\tilde\mu_0\frac{m^*}{m_0})}\rho_0(\tilde\mu_0\frac{m^*}{m_0}),
\end{equation}
where here and in the following primes indicate derivatives with respect to $h$. A similar
expression had already been derived by Luttinger \cite{Lut60}. The 
metamagnetic condition $\chi_s'(h)>0$ is then
\begin{equation}
  g''(h)\rho_0(g(h))+\rho_0'(g(h))g'(h)^2>0.
\label{metamagcond}
\end{equation}
The occurrence of metamagnetic behavior can be analyzed depending on the
functional form of $g(h)$ and $\rho_0(\epsilon)$. For a simple analysis let us
assume $h>0$ and the power law form for $g(h)=c\,h^{\alpha}$, $c>0$. The first
term in (\ref{metamagcond}) is then positive if $\alpha> 1$. For a convex density of states,
$\rho_0''(\epsilon)>0$, the second term is also positive and metamagnetic
behavior occurs as mentioned earlier. For a concave density of states, $\rho_0''(\epsilon)<0$, the
two terms in (\ref{metamagcond}) compete. If we also assume
the power law form for the density of states, $\rho_0(\epsilon)=r_0-d\,\epsilon^{\gamma}$,
(e.g. for $\rho^{\rm sem}_0$ one has $r_0=2/\pi D$
$d=r_0/2$ and $\gamma=2$) condition (\ref{metamagcond}) becomes 
\begin{equation}
  \frac{r_0}{c^{\gamma}d}\frac{\alpha-1}{\alpha(1+\gamma)-1}>h^{\alpha\gamma},
\end{equation}
Since the right hand side is positive, we can infer that for $\alpha>1$ and
$\gamma>(1-\alpha)/\alpha$ metamagnetic behavior occurs. The actual field
dependence of $g(h)$ can be calculated from the renormalized parameters and it depends on
the interaction strength. As we will see for the half filled Hubbard model and
intermediate $U$, $g(h)$ grows faster than linear with $h$, i.e. $\alpha>1$. 

In the limit of zero field the ratio of the susceptibility of the interacting
and non-interacting system has a simplified expression in terms of the
renormalized parameters,  
\begin{equation}
  \label{suscrph0}
   \frac{\chi_s}{\chi_s^0}
=g'(0)
%= \frac{m^*(0)}{m_0}\Big(\tilde\mu_0'(0)+\tilde\mu_0(0)\frac{m^*{}'(0)}{m^*(0)}\Big)  
=\frac{m^*(0)}{m_0}\tilde\mu_0'(0),
\end{equation}
for $\tilde\mu_0(0)=0$.
Comparing with the Fermi liquid expression (\ref{flsusc}) we can identify
$1/(1+F_0^a)=\tilde\mu_0'$. This quantity corresponds to the Wilson ratio $R$.
In the  general case, the field dependent enhancement due to the quasiparticle
interactions reads  
\begin{equation}
R(h)=\frac1{1+F_0^a(h)}=\Big(\tilde\mu_0'+\tilde\mu_0\frac{m^*{}'}{m^*}\Big)
\frac{\rho_0(\tilde\mu_0\frac{m^*}{m_0})}{\rho_0(h)}.
\label{flqpint}
\end{equation}

So far the considerations have been independent of our DMFT-NRG approach. In
the following section we will compare results for the magnetic susceptibility
obtained from the static expectation values of integrating the Green's
functions, with the results based on the field dependent parameters. We
determine them as described above. Alternatively they can be calculated by
other methods, such as the Gutzwiller (GW) approach, and we will make comparison as
appropriate. Results are obtained as in Ref. \onlinecite{Vol84}, where the
critical interaction for the metal insulator transition is $U^{\rm GW}_c=16
W/3\pi\approx 6.79$  for $\rho^{\rm sem}_0(\epsilon)$ with $W=4$.

\section{Results}
%\paragraph*{Results -}
\subsection{Magnetization and metamagnetic transition}
\begin{figure}[!htbp]
\centering
\includegraphics[width=0.45\textwidth]{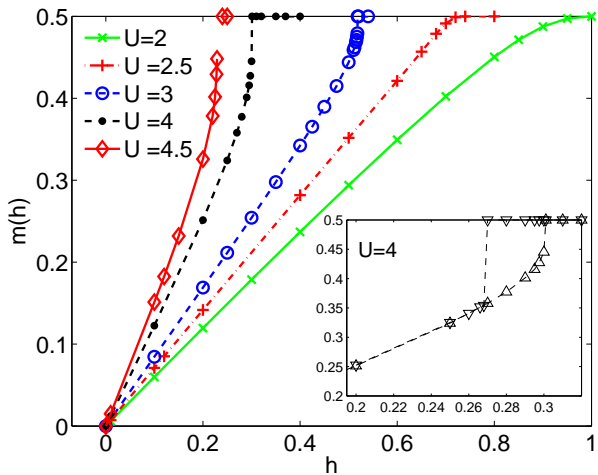}
%\includegraphics[width=0.45\textwidth]{figures_metmag/Uhplanewmag.eps}
%\vspace*{-0.5cm}
\caption{(Color online) The local magnetization $m(h)$ as a function of the
  magnetic field $h$ for different values of $U$. We can see that a
  metamagnetic curvature sets in at $U=3$. Inset: Hysteresis curve for $U=4$ (triangle up
increasing $h$, triangle down decreasing $h$).} 
%\caption{(Color online) The $U$-$h$-plane showing the different regions of
%  magnetic response: paramagnetic metal (PM) with normal magnetic response,
%  metamagnetic metal (MMM) with metamagnetic response, polarized insulator
%  (PI) with vanishing response. The background color describes the magnetization.} 
\label{maghdifU}
\end{figure}

For a first overview we present results for the magnetization $m(h)$ as a
function of field $h$ in Fig. \ref{maghdifU} for various values of $U$. 
The magnetization $m(h)$ was computed from the static NRG expectation value
(EV) for the occupation number as well as from
integrating the spectral function to the Fermi level, both of which agree very 
well. The results for $m(h)$ based on the field dependent renormalized
parameters (RP) and equation (\ref{mag2}) are also in good agreement, but not
included in the figure. 

\noindent
The plot gives a clear picture of the field strength $h_{\rm pol}$ necessary to
polarize the metal completely to $m=1/2$. For weak coupling it can be related to the
rigid band shift and a large field $h\sim D$ is needed, but for larger
interaction strength $h_{\rm  pol}$ is reduced substantially. For $U\ge 3$ a
metamagnetic curvature in the magnetization can be observed, and we see that
in the Hubbard model at zero temperature the metamagnetic transition
field \cite{hm} $h_{\rm m}$ coincides with $h_{\rm pol}$, which is not
necessarily the case for $T>0$. Laloux et al.\cite{LGK94} have compared
results from low temperature DMFT calculations with the Gutzwiller
approximation 
%(see Fig. 8 in Ref. \onlinecite{LGK94}) 
and it was found
that the occurrence of metamagnetic behavior is overestimated by the
Gutzwiller approximation (see also Fig. \ref{chihdepU4.5}).
%, e.g. for value corresponding to $U\simeq 4.2$ here
%the critical Gutzwiller field $h^{\rm GW}_m\approx 0.68 h^{\rm DMFT}_m$.

Earlier work \cite{LGK94} showed that the transition is a discontinuous
first order one at low temperature. Our results show jumps in
the magnetization curve at the transition field $h_{\rm m}$, e.g. for $U=3$ and $U=4$ in
Fig. \ref{maghdifU}, however, we can not exclude a very steep 
continuous increase which can not be resolved numerically. We have also found
hysteresis, shown for $U=4$ as an inset in Fig. \ref{maghdifU} (triangle up
increasing $h$, triangle down decreasing $h$). This suggests that the
transition is also of first order for zero temperature. 
For larger interaction $U\ge 4.5$  there exists a small field range near
$h_{\rm m}$, where  we have not found unique, well converged DMFT solutions,
so no definite statement can be made.

The half filled repulsive Hubbard model in magnetic field can be mapped to the
attractive one \cite{MRR90}, in which the chemical 
potential is related to the field in the original model, $\mu=U/2+h$. 
The attractive model has been studied by the DMFT in situations, where
superconducting order was not allowed for \cite{KMS01,CCG02}. A first order transition
from a metallic to a pairing state for fixed density was found at a
critical interaction. The occurrence of the transition can be related to the
metamagnetic transition here.
%, and where comparable the 
%conclusions agree with this work. 
A nearly polarized system corresponds to a low density limit, and to estimate
when the transition sets in, one can analyze the two-body problem in the
attractive model and calculate the critical $U_c$ for bound state
formation. For a three dimensional cubic 
lattice the result is $U_c\approx 0.659W$ \cite{MRR90}. With the given bandwidth
this corresponds to a value of $U_c\approx 2.64$, which is a reasonable
estimate for the interaction strengths, where the metamagnetic behavior 
is found here.  

\subsection{Magnetic susceptibilities and quasiparticle properties}

From the initial slope of the magnetization curves in Fig. \ref{maghdifU} we
observe an increase of the magnetic susceptibility with the interaction
strength $U$.  
%As we will see this can be understood 
%in terms of a quasiparticle picture. In such a picture the magnetic response
%is related to the behavior of the quasiparticle resonance in the field. Its
%width is characterized by $\tilde D=zD$ and the position, i.e. the shift from
%the chemical potential in zero field,  by $\tilde\mu_0$.
%Before we discuss the metamagnetic behavior let us understand the magnetic
%response in zero field for increasing interaction in terms of these
%quasiparticle parameters. 
This increase can also be seen in the following Fig. \ref{chiUdep} where we
show the ratio of zero field susceptibility to the non-interacting value
$\chi_s^0$ as function of $U$ deduced from differentiating the EV for $m(h)$
in the limit $h\to 0$.  

\begin{figure}[!htbp]
\centering
\includegraphics[width=0.45\textwidth]{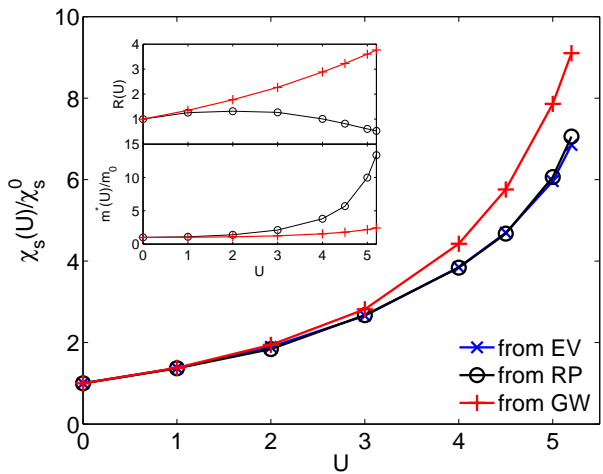}
%\vspace*{-0.5cm}
\caption{(Color online) The $U$-dependence of the magnetic susceptibility
  $\chi_s$. We compare results deduced from the EV of $m(h)$ with ones
  obtained from the RP and from the Gutzwiller (GW) approximation. The inset shows
  the effective mass $m^*(U)/m_0$ and the Wilson ratio $R(U)$ as a function of $U$.}   
\label{chiUdep}
\end{figure}
\noindent
For comparison we have also included the susceptibility calculated from
equation (\ref{suscrph0}) with the renormalized parameters (RP) and their
derivatives, as well as the results obtained from the Gutzwiller (GW) approximation. 
EV and RP results agree very well, confirming the applicability of Fermi liquid
results in this metallic regime. The GW results follow a similar trend but
overestimate the value for the susceptibility, which becomes more pronounced
for larger $U$. 

The inset plot shows the $U$-dependence of
the effective mass and the Wilson ratio. In terms of Fermi liquid theory and the expression
(\ref{flsusc}) the increase of $\chi_s$ with $U$ can be understood by the 
behavior of the effective mass and the progressive localization tendency,
which brings out more the spin degrees of freedom of the electrons. We can
see, however, that the effective mass ratio is larger than that of the
magnetic susceptibility. This difference can be attributed to the factor
$R=\tilde\mu_0'=[1+F_0^a]^{-1}$, which is due to the quasiparticle
interaction. This factor is larger than one for smaller values of $U$, but
decreases to values below one for stronger interaction. This indicates a sign
change of the parameter $F_0^a$ from negative to positive. The comparison of
the corresponding quantities calculated in the GW approximation shows
a qualitatively similar behavior for both $m^*/m_0$ and $R$, 
when $U$ is small. For larger values of $U$ in Fig. \ref{chiUdep}, however,
the effective mass enhancement in the GW approach, $m^*/m_0=1-(U/U_c^{\rm
  GW})^2$, is much smaller and $R$ increases with $U$ in contrast to the DMFT result.
%Closer to the metal insulator transition the effective mass enhancement will
%become the main drive for the increasing susceptibility.
%the Gutzwiller approximation $F_0^a$ tends to a constant value. In the DMFT
%calculations here, we have not studied this regime in sufficient detail to
%make definite predictions. 

%, which is less than one here. We can calculate the value $\tilde
%U$, which is related to the quasiparticle interaction and it is found to
%decrease in with increasing $U$. 

\begin{figure}[!htbp]
\centering
\includegraphics[width=0.45\textwidth]{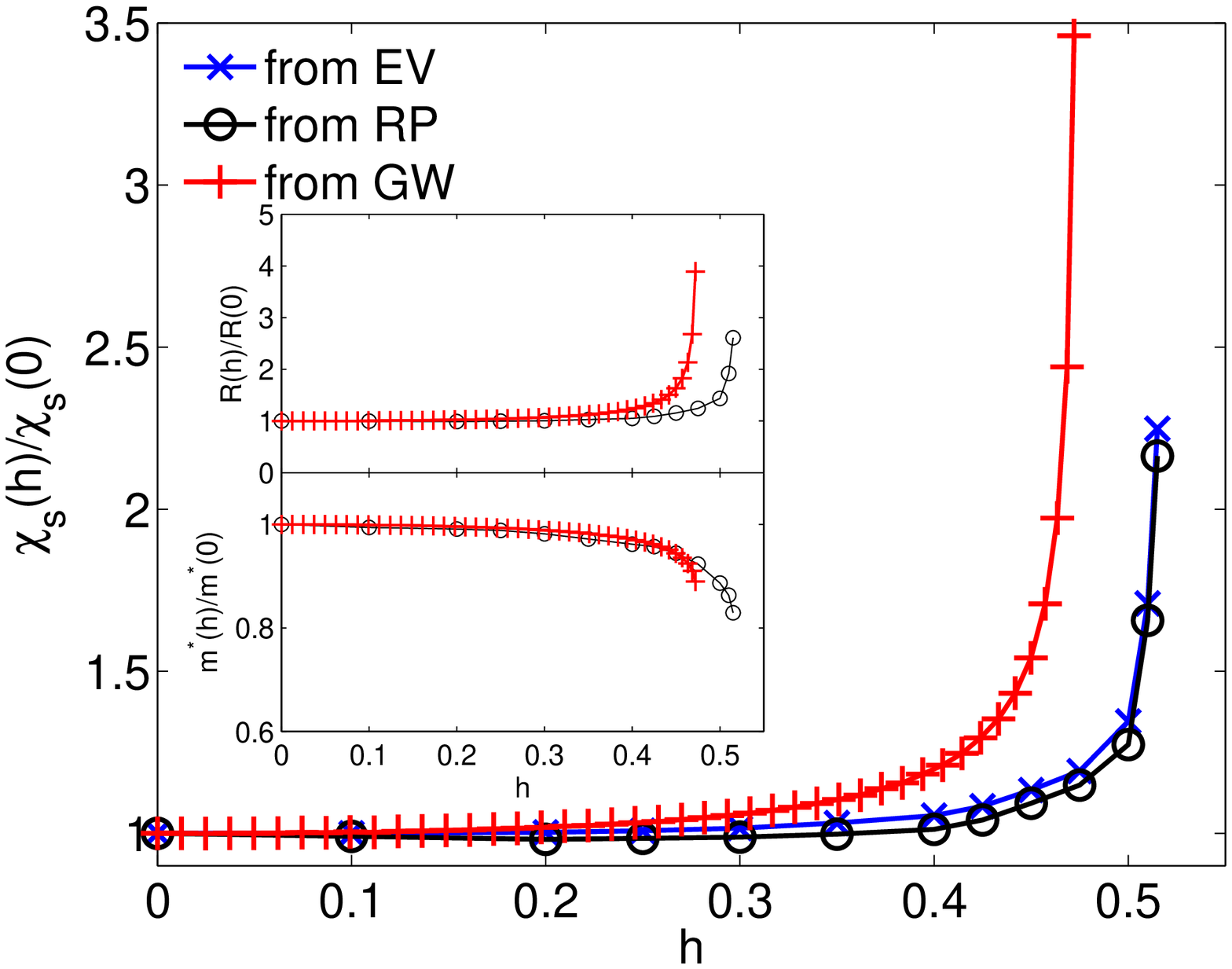}
\includegraphics[width=0.45\textwidth]{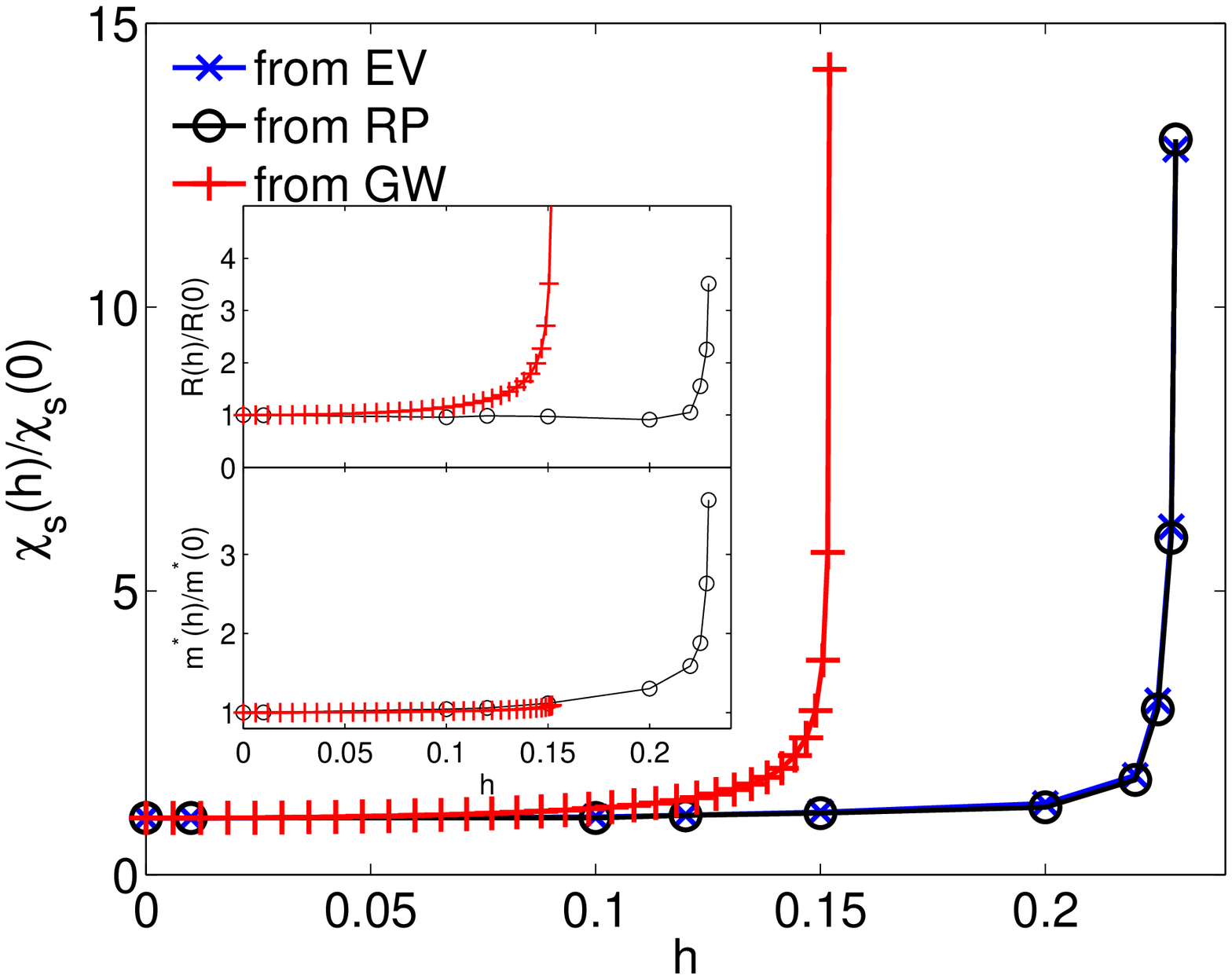}
%\vspace*{-0.5cm}
\caption{(Color online) The $h$-dependence of the ratio of the finite and zero
  field magnetic susceptibility  $\chi_s$ for $U=3$ (upper panel) and $U=4.5$
  (lower panel). We compare results deduced from the EV for $m(h)$
  with ones obtained from the RP and the ones from the GW approach. 
  The inset shows the ratio of finite and zero field effective mass
  $m^*(h)/m_0(0)$ and the Wilson ratio $R(h)/R(0)$ as a function of $h$.}   
\label{chihdepU4.5}
\end{figure}

%cf Fermi liquid
We return the finite field response and focus on the
metamagnetic behavior which is found for intermediate values of $U$.
%in the regime $3W/4<U<U_c$, where $U_c$
%is the critical interaction for the Mott-transition.
Results for the ratio of the magnetic susceptibility in finite and zero field
deduced from differentiating the magnetization (EV) are compared to 
the ones obtained from the quasiparticle parameters (RP) and equation
(\ref{suscrp}). For completeness, we have also included results from the GW approximation. This
is shown in Fig. \ref{chihdepU4.5} for $U=3$ in the upper panel and $U=4.5$ in
the lower panel.   

\noindent
We can see that also in finite field the results for the susceptibility  
calculated from the EV for $m(h)$ and the field dependent RP agree fairly well
with a deviation of less than 5$\%$. For the case $U=3$ (upper panel) the
results for $\chi(h)$ based on the field 
dependent RP are always smaller. In both cases we find first a period where
the susceptibility is nearly constant, but then starts to increase rapidly as
$h$ approaches $h_{\rm   m}$. For $U=3$ the values obtained from the RP
initially decrease slightly with the field, 
which is however incorrect, and comes about through numerical inaccuracies when
determining the parameters and the numerical differentiation. As $h_{\rm
  m}=h_{\rm   pol}$  the magnetic susceptibility is zero for $h>h_{\rm m}$. At
finite temperature a susceptibility maximum is expected. The results for
$\chi_s$ from the GW approximation show generally a similar trend, but as
mentioned earlier the metamagnetic behavior sets in at lower field strengths.

A difference in the behavior between the two cases is visible in the two
insets where the ratios of field dependent effective masses to their zero field
values and the field dependent Wilson ratios $R(h)/R(0)$ are plotted. For the
$U=3$ case the effective mass decreases with the field which is typical
behavior in the weak coupling regime. It can be understood by RPA
approximations where spin fluctuations, which give an effective mass
enhancement, are suppressed in finite field. The metamagnetic increase of the
susceptibility, however, can not be explained by this. In terms of Fermi
liquid theory it is related to the magnetic field dependence of the
quasiparticle  
interaction rather than the localization tendency encoded in the effective
mass. $R(h)/R(0)$ indeed is increasing sharply close to $h_{\rm m}$. In equation
(\ref{flqpint}) we have two competing terms for this enhancement
factor,  $m^*{}'/m^*<0$, but one finds $\tilde \mu_0'>|\tilde \mu_0\,m^*{}'/m^*|$ 
which leads to the observed enhancement. The drive for the metamagnetic
behavior is therefore due to the shift of the quasiparticle band  
from the Fermi level with increasing field. This contrasts to the weak
coupling situation, such as $U=2$, where $R(h)$ decreases with the field strength and no
metamagnetic response is observed.

The effective mass in the case of $U=4.5$ (lower panel in Fig.
\ref{chihdepU4.5}) shows  different behavior. %from the$U=3$ case. 
We can see a sharp increase with the field. However, the magnitude
the ratio $m^*/m_0$  increases is less than that of the susceptibility. The
difference again can be related to the Fermi liquid factor $R=1/[1+F_0^a]$,
which is larger than one and increasing with $h$ as can be seen in the inset
of the lower panel in Fig. 
\ref{chihdepU4.5}. In this case the second term in equation (\ref{flqpint}) is
positive and the first term negative, but $|\tilde  \mu_0'|<|\tilde \mu_0\,m^*{}'/m^*|$.  
The results from GW approach for the effective mass and $R$ are in line with the DMFT
calculations for the case $U=3$, however, for $U=4.5$, the GW result for
$m^*{}'/m^*$ only increases very little with the field, whereas $R(h)$ increases
sharply to yield the metamagnetic response.

For larger interactions than the ones discussed here ($5<U<U_c$), 
one can encounter difficulties to reach convergency in the DMFT calculations with
finite field as discussed in earlier work\cite{BH07b}. The results indicate,
however, that there is a strong field dependent enhancement of the effective
mass which is the main  drive for the metamagnetic response. The ratio
$R(h)/R(0)$ varies little with $h$ or even decrease for larger fields. Such a
behavior is also found within the GW approach for larger $U$ near the metal
insulator transition.

\subsection{Spectral functions}
\begin{figure}[!htbp]
\centering
\includegraphics[width=0.45\textwidth]{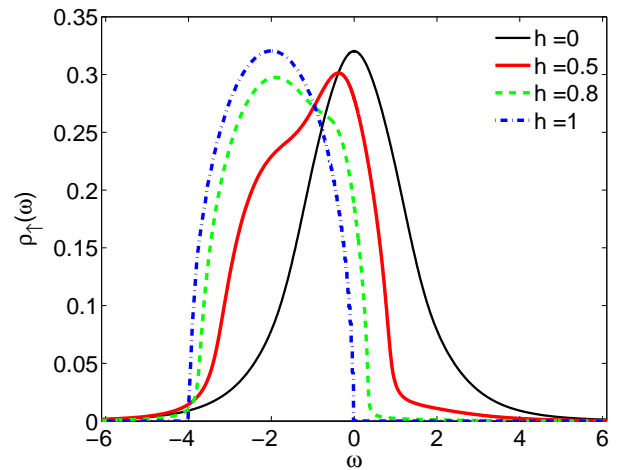}
%\vspace*{-0.5cm}
\caption{(Color online) The majority spin density of states for $U=2$ and various field
  strengths in comparison.}    
\label{dosU2varh}
\end{figure}

The behavior of the quasiparticle band can be seen directly in the local
spectral function. For the cases with smaller coupling the field dependent
response shows a continuous shift of spectral weight to
lower energies for the majority spin (see Fig. \ref{dosU2varh} for $U=2$).

\begin{figure}[!htbp]
\centering
\includegraphics[width=0.45\textwidth]{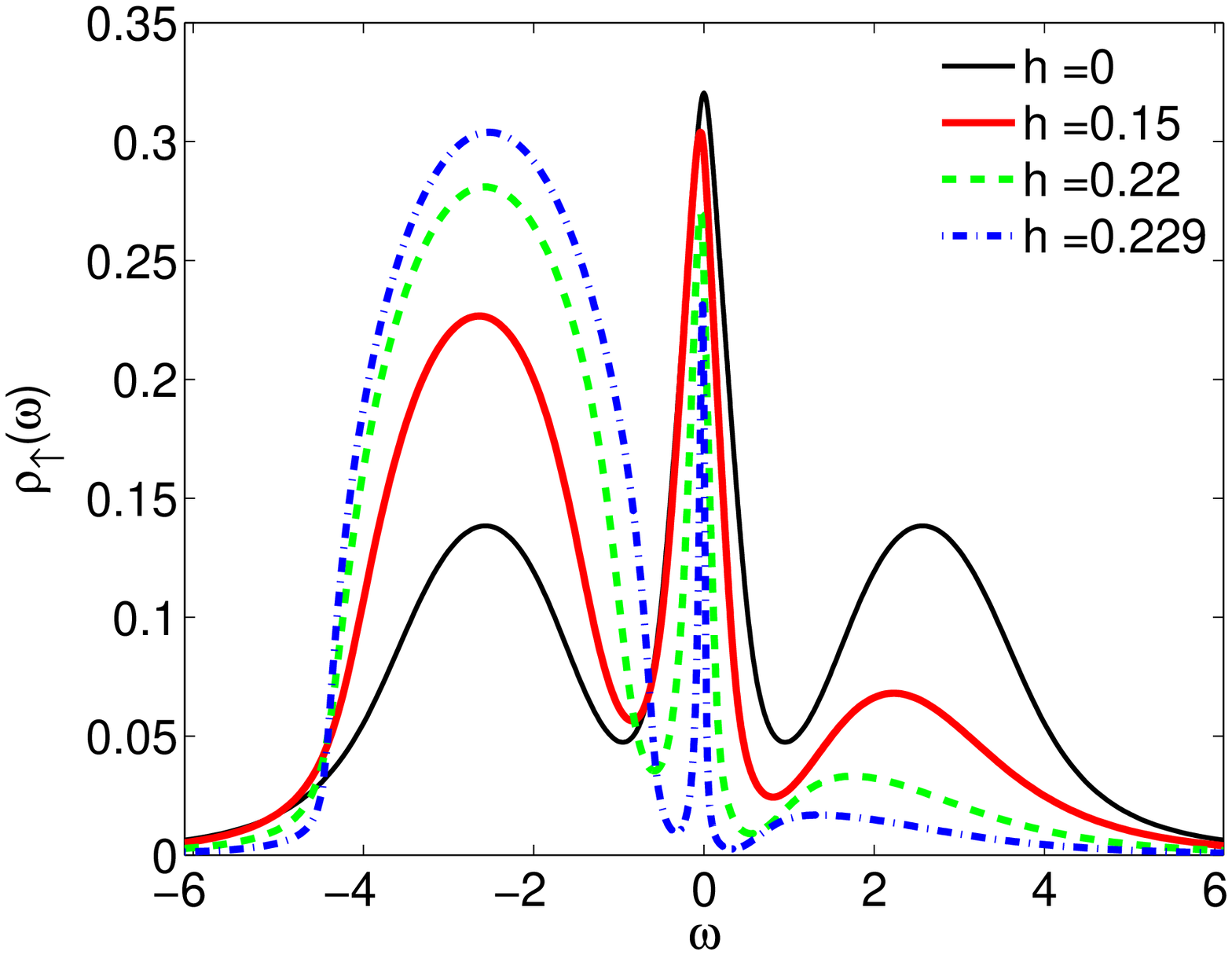}
\includegraphics[width=0.45\textwidth]{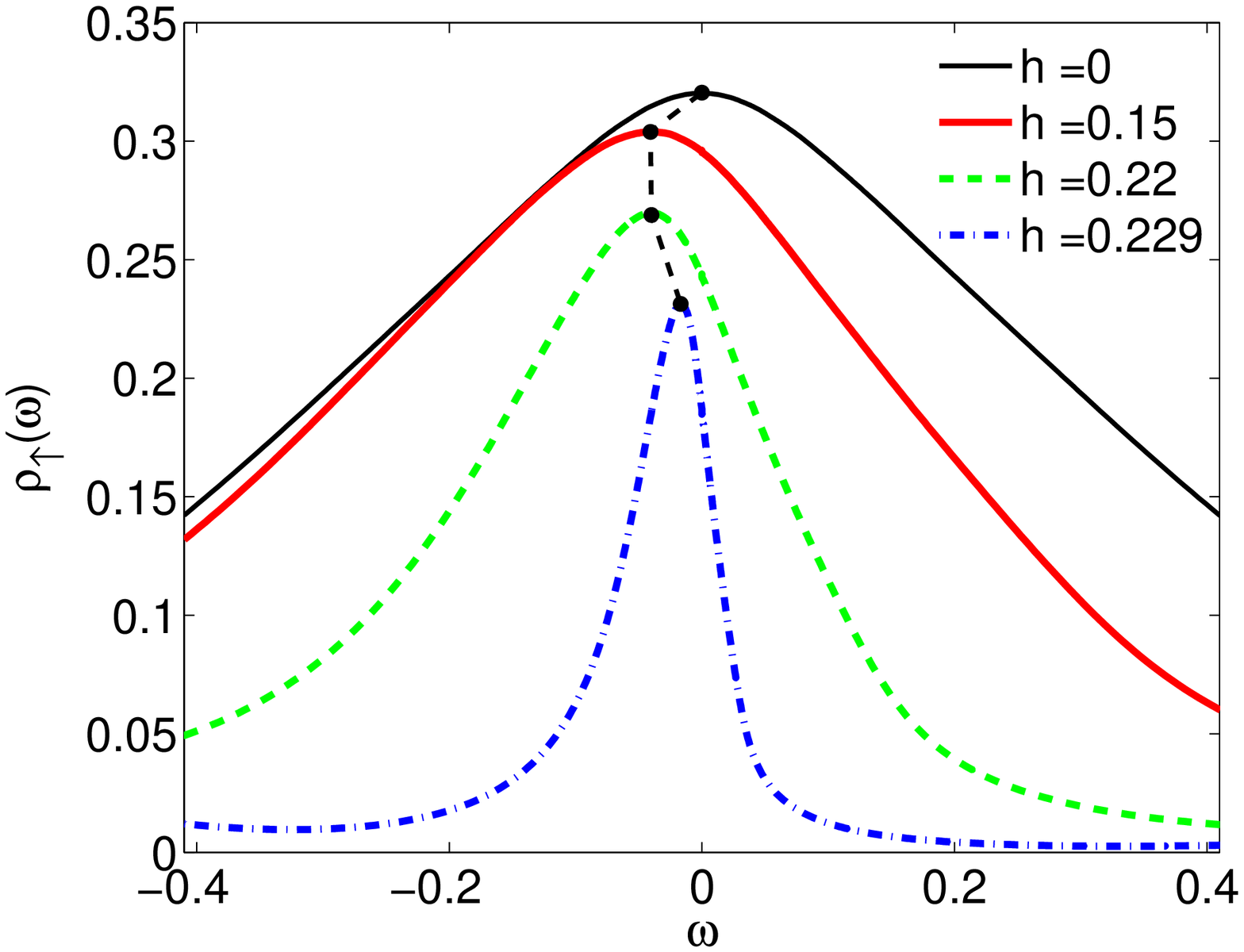}
%\vspace*{-0.5cm}
\caption{(Color online) The majority spin density of states for $U=4.5$ and various field
  strengths in comparison: upper panel full frequency range, lower panel low
  frequency behavior.}    
\label{dosU4.5varh}
\end{figure}

\noindent
Note that the minority spin density of states $\rho_{\downarrow}(\omega)$
is given by $\rho_{\uparrow}(-\omega)$ at half filling.
To illustrate  the behavior of the quasiparticle peak for the stronger
interacting case with $U=4.5$ in more detail, we plot the local spectral
function for the majority spin $\rho_{\uparrow}(\omega)$  in
Fig. \ref{dosU4.5varh}.     

\noindent
In the upper panel we can see how the lower Hubbard peak in the spectral
density acquires weight when the field and thence magnetization is
increased whilst the upper Hubbard peak loses spectral
weight. The behavior at low energy is seen more clearly in the lower
panel. At first sight the overall picture is reminiscent of the particle hole
symmetric Anderson impurity model in the Kondo regime in magnetic field
\cite{HBK06} as far as the high energy behavior is concerned. The
quasiparticle resonance in the locally correlated system broadens and departs
from the Fermi level. This behavior occurs in an analogous
fashion in the weak coupling regime of the Hubbard model with
$\tilde\mu_0'(h)>0$.
%, although there is not such a strongly pronounced quasiparticle peak there. 
In the strongly correlated case, however, we find a
significant narrowing of the quasiparticle peak in the field, which is accompanied
by the field induced metal insulator transition and metamagnetic
behavior. The quasiparticle resonance first departs from the Fermi
energy, but for larger fields is driven back to it. These features are visible
in the field dependence of the renormalized parameter $\tilde\mu_0$ with
$\tilde\mu_0'<0$ as discussed above. 
%Such a behavior can only be described in
%an approach which properly accounts for the strong correlation effects.  

\section{Relation to experiments and conclusions}

It is of interest to see, whether the described behavior bears any resemblance
with what is observed experimentally in strongly correlated itinerant electron
system.  
Metamagnetic behavior is observed, for instance,  
%in antiferromagnetic insulators FeCl${}_2$,DyPO${}_4$,\cite{LTBW90}
in the heavy fermion compounds CeRu${}_2$Si${}_2$
 \cite{PLPHLTF90,FHRAK02}, UPt${}_3$ \cite{MTVFPAK90} or
 Sr${}_3$Ru${}_2$O${}_7$  \cite{LTBW90,GPSCJLIMM01,FHRAK02,PTKSIM05} and  
the Co-based metallic compounds such as
Y(Co${}_{1-x}$Al${}_x$)${}_2$,\cite{SGYF90,GKSMFM94} 
sometimes called nearly ferromagnetic metals.
%also the high-$T_c$ compound La${}_2$CuO${}_4$ \cite{TTPPKJGCBA88,CTF89}. 
The microscopic origin for the occurrence of the effect in these compounds can
be manifold, and is sometimes still controversial. In many cases
antiferromagnetic exchange is thought be important and the system's
closeness to a magnetic instability.

For generic features, we attempt to compare our microscopic Fermi liquid
description with experimental studies of itinerant metamagnetic behavior in
heavy fermion compounds. It is important, however, to be aware that our
results based on the paramagnetic solutions of the half filled single band
Hubbard model are not appropriate to make quantitative predictions for those
complex systems. Organic conductors are thought behave
like simple Mott-Hubbard systems and  have been
shown to display a magnetic field induced localization transition with
hysteresis by resistance measurements.\cite{KIMK04} The author is, however,
not aware of any published field dependent magnetization or specific heat data to compare to.

In materials  such as CeRu${}_2$Si${}_2$, UPt${}_3$ or
 Sr${}_3$Ru${}_2$O${}_7$ the magnetic field  
dependence of the linear specific heat coefficient $\gamma$ was measured near
the metamagnetic transition \cite{PLPHLTF90,MTVFPAK90,FHRAK02,PTKSIM05}. It is
worth noting that, as can be shown from a thermodynamic identity, the field
dependence of $\gamma$ can also be extracted from $T^2$-coefficient of the
magnetization \cite{PLPHLTF90}. In the experiments $\gamma$
increases with the magnetic field and possesses a maximum at the metamagnetic transition
$h=h_{\rm m}$. This is comparable with the Fermi liquid results for stronger 
coupling, e.g. the case $U=4.5$ (Fig. \ref{chihdepU4.5} lower panel), where
the effective mass increases with the magnetic field. In the case of
CeRu${}_2$Si${}_2$ \cite{FHRAK02} one can see that the susceptibility increase with the 
magnetic field is up to about 8.5 times the zero field value, whereas in the
same regime the specific heat coefficient only shows an enhancement of
1.6. In our Fermi liquid interpretation this signals that the quasiparticle
interaction plays an important role in the susceptibility enhancement and the metamagnetic
behavior. The relevance of this has been emphasized in the recent experimental
work on Yb${}_3$Pt${}_4$.\cite{BSKJYGA08pre}
A more careful quantitative comparison would be possible based on
the periodic Anderson model, for instance. The presented approach can be extended to this
situation, but also other techniques are available
\cite{MN01,SI96,Ono98,EG97}.   
To summarize, we have analyzed the metamagnetic response of the half
filled Hubbard model in terms of renormalized quasiparticle parameters and
Fermi liquid theory. The renormalized parameters can be calculated accurately with methods
based on the NRG, and they have a clear physical meaning. It is shown that the
field dependent metamagnetic behavior  
can have part of its origin in field induced effective mass enhancements, but
is not fully explained by this. This is most clearly pointed out in Fig.
\ref{chihdepU4.5}, where metamagnetic behavior for smaller $U$ is accompanied
by an effective mass reduction in the field, whereas for larger interaction the opposite
is the case. The comparison with results obtained from the Gutzwiller
approximation gives similar trends, but shows quantitative deviations.
The hypothesis that the metamagnetic behavior in itinerant
systems is always driven by field induced mass enhancement is therefore found
to be not valid. In the intermediate coupling regime it is also shown 
that the effective mass enhancement alone is not sufficient to explain the
metamagnetic enhancement and based on Fermi liquid theory arguments the
quasiparticle interaction has to account for the difference. As a generic
feature there the corresponding term described by the Wilson ratio $R$
increases near the metamagnetic 
transition. The opposite happens in the weak (no metamagnetic response) and
strong coupling situation. The observation that only a part of the
susceptibility enhancement is based on the effective mass is found to be
qualitatively in agreement with experimental observations in heavy fermion
systems.

\par
%\bigskip
\noindent{\bf Acknowledgment}\par
\noindent
I wish to thank K. Held, A.C. Hewson, P. Jakubczyk, W. Metzner, A. Toschi, D. Vollhardt,
and H. Yamase for helpful discussions, W. Koller and D. Meyer for their
earlier contributions to the development of the NRG programs, and A. Toschi
for critically reading the manuscript. I would like to acknowledge many fruitful 
discussion with A.C. Hewson during early stages of this work and thank the
Gottlieb Daimler and Karl Benz Foundation, the German Academic Exchange
Service (DAAD) and the EPSRC for financial support during this
period.

\bibliography{artikel,biblio1,footnote}

\begin{thebibliography}{46}
\bibitem[{hm()}]{hm}
\bibinfo{note}{The metamagnetic transition field is the field where the
  susceptibility is maximal.}
\end{thebibliography}

\end{document}